\documentclass[a4paper,nocompress]{spie}  
\usepackage{amsmath,amsfonts,amssymb,units}
\usepackage{gensymb,textcomp}
\usepackage{graphicx}
\usepackage[colorlinks=true, allcolors=blue]{hyperref}
\usepackage[normalem]{ulem}
\usepackage[toc,page]{appendix}
\usepackage[section]{placeins}
\usepackage{multirow}
\usepackage[colorinlistoftodos]{todonotes} % see documentation: https://ctan.mirror.norbert-ruehl.de/macros/latex/contrib/todonotes/todonotes.pdf

\usepackage{soul} % use \st{text} to strike out text

\newcommand{\FBG}{\text{FBG}}

\title{Athermal package for OH suppression filters in astronomy\\ part 1: design}
\author[a,b]{Carlos Rordriguez Alvarez}
\author[a]{Aashia Rahman}
\author[a]{Hakan \"{O}nel}
\author[a]{Frank Dionies}
\author[a]{Jens Paschke}
\author[a]{Svend-Marian Bauer}
\affil[a]{Leibniz-Institut f\"{u}r Astrophysik Potsdam (AIP), An der Sternwarte 16, 14482 Potsdam, Germany}
\affil[b]{Technische Universit\"{a}t Berlin, Fakult\"{a}t V - Institut f\"{u}r Mechanik, FG Strukturmechanik und Strukturberechnung, Sekr. C 8-3, Geb. M, Stra\ss{}e des 17. Juni 135, 10623 Berlin, Germany}

\authorinfo{Further author information: (Send correspondence to C.R.A. and A.R.)\\Carlos Rordriguez Alvarez (C.R.A.): E-mail: \url{crodriguez@aip.de} \\ Aashia Rahman (A.R.): E-mail: \url{arahman@aip.de}}

\begin{document} 
\maketitle

\begin{abstract}
We present the design of an athermal package for fiber Bragg grating (FBG) 
filters fabricated at our Institute for use in ground-based near-infrared (NIR) telescopes. Aperiodic multichannel FBG filters combined with
photonic lanterns can effectively filter out extremely bright atmospheric hydroxyl (OH) emission lines that severely hinder ground-based NIR observations. While FBGs have the capability of filtering specific wavelengths with high precision, due to their sensitivity to temperature variations, the success in their performance as OH suppression filters depends on a suitable athermal package 
that can maintain the deviations of the FBG wavelengths from that of the OH emission 
lines within sub-picometer accuracy over a temperature range of about \unit[40]{K}
(i.e. $\unit[263]{K}$ to $\unit[303]{K}$). 
We aim to develop an athermal package over the aforementioned temperature range for an optical fiber
consisting of multichannel FBGs for a maximum filter length of $\unit[110]{mm}$.
In this work, we demonstrate the complete design methodology of such a package.
First, we developed a custom-built test rig to study a wide range of critical 
physical properties of the fiber, such as strain and temperature sensitivities, elastic modulus, 
optimum fiber pre-tension, and adhesion performance.
Next, we used these data to confirm the athermal response of an FBG bonded on the test rig 
from room temperature to $\unit[313]{K}$. Based on this study, we developed 
a computer-aided design (CAD) model of the package and analyzed its athermal characteristics with 
a suitable selection of materials and their nominal dimensions using finite element analysis (FEA). We finally discuss the novel aspects of the design to achieve high-precision thermal stabilization of these filters in the temperature range of interest.
Our design is scalable to longer filter lengths and can also be used for other in-fiber devices for
example, FBG-based frequency combs for their stable operations in astronomical applications.

\end{abstract}

\section{INTRODUCTION}
Hydroxyl (OH) radicals formed by the reaction of atomic hydrogen and ozone at an altitude of approximately \unit[87]{km} in the Earth's atmosphere, emit between $\unit[0.61]{\micro m}$ and $\unit[2.62]{\micro m}$ \cite{Rousselot}. These emission lines are particularly intense in the near-infrared (NIR) spectra, posing significant challenges for ground-based astronomical observations.  Efficient removal of these lines from astronomical spectra is essential for ground-based astronomy. This becomes even more critical for the upcoming Extremely Large Telescope (ELT), which will focus extensively on NIR observations.

\par FBGs offer a promising solution for filtering out OH lines in the NIR range. Astronomical instruments such as GNOSIS and PRAXIS have successfully demonstrated the potential of FBGs for this purpose \cite{Ellis1,Ellis2}. However, FBGs are highly sensitive to temperature fluctuations, with a typical temperature sensitivity of around $\unit[10]{pm/K}$. For a spectrograph with a resolving power of $R =50\,000$, the spectral resolution ($\Delta\lambda$) at $\unit[1\,550]{nm}$ is $\unit[31]{pm}$, necessitating a temperature stability for an FBG to be within $\unit[3]{K}$ over the operational temperature range. Accommodating other factors (e.g., accuracy of the filter design, and fabrication), the thermal stabilization of an FBG OH filter must aim for temperature stability within $\pm \unit[0.1]{K}$, ensuring wavelength stability of $\le \unit[1]{pm}$. Achieving the tight tolerance in thermal stability is critical to ensure that the interline continuum \cite{Ellis0} is not masked by the filter's profile. Therefore, a suitable package to thermally stabilize the FBG filters within sub-picometer precision and accuracy becomes crucial. 

\par Commercially available athermal packages for FBGs cannot be readily adopted, as these packages typically isolate the grating and contain inbuilt single notch FBGs unsuitable for astronomical applications. Although a few commercial solutions offer customized packaging options for FBGs, the filters must often be procured from the same companies. Manufacturing FBG filters for OH suppression is complex and requires an advanced FBG fabrication setup \cite{Skaar, Yu}. At Leibniz Institute for Astrophysics Potsdam (AIP), we fabricate these filter lines using an advanced Talbot interferometry fabrication facilities while exploring novel methods to enhance the reproducibility of FBG fabrication.\cite{Rahman_Opex,Rodriguez24,Rahman23} 
To ensure high precision filter stabilization at ambient temperatures ($\unit[263]{K}$ to $\unit[303]{K}$), we have designed and manufactured a self-compensating package for these OH filters.

\par In this work, we present a detailed description of the design process for such an athermal unit. Section~\ref{sec:theory} discusses the basic theory of an athermal package, emphasizing the critical aspects to consider for the design. Section~\ref{sec:method} outlines the design methodology, followed by the complete experimental realization in Section~\ref{sec:experimentalproc}. Finally, in Section~\ref{sec:discussions} we discuss the novel aspects of our design.

\section{Basic theory of an athermal package}
\label{sec:theory}  % \label{} allows reference to this section
Considering the extensive utilization of FBGs in  telecommunication and sensing applications \cite{Othonos, Kashyap, Allwood} over the past several decades, a diverse range of packaging methods and temperature compensation techniques has emerged to enhance their efficacy as devices. Kuang \emph{et al.} \cite{Kuang} provides a comprehensive overview of three primary packaging methods for FBG devices, along with their associated temperature compensation techniques: a) fully bonded FBG, b) pre-stretched FBG with double-end fixed, and c) metallic packaging. Notably, method b) stands out for its ability to eliminate chirp along the length of the FBG while ensuring straightforward operability. 
In this paper, our objective is to develop an athermal packaging solution tailored for multichannel FBG filters for astronomical applications. The filter length can be as long as $\unit[110]{mm}$; no chirp or non-uniform strain must act along the length of the filter. For astronomical applications, the temperature range is not crucial, but for a stable and effective filter operation, the deviation in wavelength should be within sub-picometer range. We focus on developing an athermal packaging solution that meets this stringent demand.

\par
\begin{figure}[ht]
\centering
\includegraphics[width=1.0\textwidth]{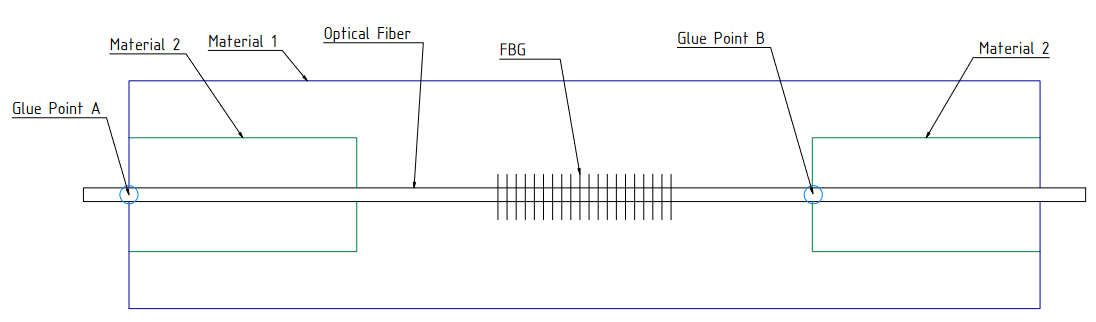} 
\caption[concept]{Schematic diagram of a dual-material athermal package with a pre-stretched FBG bonded at both ends of the structure. Material 1: Material with low CTE, Material 2: Material with high CTE, Glue Point, A and B, are the affixed/anchored points.}
\label{fig:concept}
\end{figure}

 We choose a dual-material hollow tubular design wherein the optical fiber consisting of the FBG filter is bonded at both ends of the tubular structure using adhesive.  The core principle of this athermal packaging approach lies in the structural expansion or contraction of the tubular framework, thereby applying or relieving strain on the filter to counteract wavelength shifts induced by temperature fluctuations. Thus, the coefficient of thermal expansion (CTE) and thermal conductivity of the materials and the filter's response to strain and temperature emerge as critical parameters influencing the design of the athermal package. 
Figure \ref{fig:concept} shows a schematic of a fundamental dual-material structure with a pre-stretched FBG affixed at both ends of the structure. Depending on the specific temperature range of the application, further enhancements can be made by transitioning to a multi-material structure.  Lachance \emph{et al.} \cite{Lachance} provides insightful theoretical groundwork on the thermal behavior of such athermal packaging configurations.

\par In this section, first, we discuss the theory of temperature and strain effects \cite{Magne} in FBG, and then, we mention the critical equations \cite{Lachance} to achieve the temperature compensation in the filter package. 
\subsection {Strain and temperature sensitivities of an FBG}
An FBG is a periodic microstructure in the core of a single-mode optical fiber that enables a precise wavelength filtering of the light transmitted through the fiber. This periodic microstructure is created by modulating the core refractive index by a high power laser beam to which the optical fiber is photosensitive. The modulation in the core refractive index couples light between the forward propagating and the backward propagating modes. The Bragg wavelength of a grating, $\lambda_\text{B}$, is the wavelength at which constructive interference occurs between the counter-propagating modes within the grating structure. $\lambda_\text{B}$ is given by the Bragg phase-matching condition
\begin{equation}\label{eq:1}
\lambda_\text{B}= 2n_{\text{eff}} \Lambda,
\end{equation}
where $n_{\text{eff}}$ is the effective refractive index of the core of the fiber, and
$\Lambda$ is the fringe spacing of the grating, i.e., the periodicity of the refractive index modulation. The relative change in the Bragg wavelength is given as 
\begin{equation}\label{eq:2}
\frac{\Delta\lambda_\text{B}}{\lambda_\text{B}}= \frac{\Delta n_{\text{eff}}}{n_{\text{eff}}} + \frac{\Delta\Lambda}{\Lambda}.
\end{equation}
Both $n_{\text{eff}}$ and $\Lambda$ depend on temperature and strain and the combined effect on the FBG will shift the wavelength as given as 
\begin{equation}\label{eq:3}
\frac{\Delta\lambda_\text{B}}{\lambda_\text{B}} = \epsilon_{x}-\frac{n_\text{core}^2}{2}\left[{\epsilon_{r}}\left(\text{P}_{11} + \text{P}_{12}\right) + {\text{P}_{12}} {\epsilon_{x}}\right]+ \left(\alpha_{f} + \zeta\right)\Delta {T},
\end{equation} where the $\epsilon_{x}$ is the axial strain, $\epsilon_{r}$ is the radial strain, $n_\text{core}$ is the refractive index of the fiber core in which the FBG is inscribed, $\alpha_{f}$ and $\zeta$ are the thermal expansion coefficient and thermo-optic coefficient of the fiber material, respectively. In Eq.~(\ref{eq:3}), the first term denotes the fringe spacing changes due to $\epsilon_{x}$, the second term associated with ${n_\text{core}}$ refers to photoelastic changes in the refractive index of the fiber core, and the third and fourth term refer to change of shape and refractive index of the fiber material with change in temperature, $\Delta {T}$. 
When an FBG is strained axially, $\epsilon_{r}$ is determined by Poisson's law as follows:
\begin{equation}\label{eq:4}
\epsilon_{r} = -\text{v}_{s}\epsilon_{x},
\end{equation}

For a germanosilicate fiber, $n_\text{core} \approx 1.46$, $\alpha_{f} \approx 5 \times 10^{-7}$ K$^{-1}$  and $\zeta \approx \unit[7 \times 10^{-6}]{K^{-1}}$. Considering, $\text{v}_{s} \approx 0.17$ for silica, we get, photo-elastic constant, \mbox{$p_e=({n_{\text{core}}^2}/2)[{\text{P}_{12}} - \text{v}_{s}({\text{P}_{11}} + {\text{P}_{12}})]$} $\approx 0.22$.
Now, Eq.~(\ref{eq:3}) can be simplified to 
\begin{equation}\label{eq:5}
\frac{\Delta\lambda_\text{B}}{\lambda_\text{B}}= (1-\text{p}_{e})\epsilon_{x}+ (\alpha_{f} + \zeta)\Delta {T},
\end{equation} 
 
Equation~(\ref{eq:5}) leads to an FBG strain sensitivity $\approx \unit[1.2]{pm}/{\mu \epsilon}$ and a temperature sensitivity $\approx \unit[10]{pm/K}$ at $\unit[1\,550]{nm}$. 
The strain and temperature responses of an FBG are influenced not only by fiber properties such as effective refractive index, Young's Modulus, and Poisson's ratio but also by factors related to FBG inscription methods\cite{Rahman1,Rahman2}. Therefore, it becomes imperative to accurately assess the strain and temperature sensitivities of the specific FBG intended for use in the device.

\subsection {Athermalization in a packaged FBG}
To achieve temperature compensation in an athermal package, it is important to consider  the behavior of an FBG in the athermal unit, which can be described \cite{Lachance} by Eq.~(\ref{eq:6}):
\begin{equation}\label{eq:6}
\frac{\Delta\lambda_\text{B}}{\lambda_\text{B}}= \xi\Delta{T}+\alpha_a\Delta{}T-\text{p}_{e}(\alpha_a-\alpha_f)\Delta{}T,
\end{equation}
The first term on the right-hand side of the Eq.~(\ref{eq:6}) expresses the thermo-optic effect on the refractive index in the optical fiber. The second term represents the CTE of the athermal package ($\alpha_{a}$) characterizing the thermal behavior of the distance between the both anchor points $A$ and $B$, as shown in Figure~\ref{fig:concept}. 
The last term represents the effect of the index refraction due to mechanical stress, which must be taken into account because the optical fiber  must be mounted with a certain preload so that this tension is released when the temperature increases.
Athermalization is reached when the right-hand side of Eq.~(\ref{eq:6}) is equal to $0$, for example:

\begin{equation}\label{eq:7}
\alpha_a=-\frac{\alpha_f\text{p}_{e}+\xi}{1-\text{p}_{e}}
\end{equation}

\section{Design methodology}
\label{sec:method}
The design of an athermal package for precise compensation of temperature variation in the FBG filters depends on several critical aspects: a) the adhesive material, b) the pre-strain applied to the fiber, and c) the strain and temperature sensitivities of the FBG. This section presents an overview of our methodology for designing an athermal package for FBG OH-suppression filters or any FBG-based in-fiber device, such as a Fiber Fabry-P{\'e}rot interferometer, that needs high precision in temperature stabilization.
Our methodology includes the following steps: steps 1 to 3 - build the test rig, validate it, and assess its stability; step 4 - determine the FBG properties for temperature compensation; step 5 - cross-verify the results obtained in step 4; and finally, step 6 - design the athermal package using CAD and FEA.
\begin{enumerate}
        \item Development and validation of a custom test rig: In this step, we developed a custom-built test rig with fiber holders (with V-grooves). The fiber consisting of an FBG is anchored at its two ends into the test rig. To validate the test rig, we measured a known parameter, Young's Modulus of the optical fiber. We present the details of the experimental setup in the next section.
        \item Testing of adhesion stability: A cyanoacrylate glue or a superglue is used to bond the fiber on the V-grooves. We conducted the stability tests for the glue and the stage over 24~hours. This stability test is essential in choosing the glue type and the pre-strain amount that should be applied to the fiber to avoid any slippage/breakage at the gluing junctions.
         \item Selection of pre-strain: The selection of the pre-strain depends majorly on three important aspects: a) the stability of the glue, b) the allowed pre-strain during the FBG fabrication process, c) and the range of temperatures for which the athermal package is designed to compensate for temperature variation.
        \item Determination of strain and temperature sensitivities of FBG: The strain and temperature sensitivities of an FBG depend on the grating inscription process and the type of photosensitive fiber. Therefore, it is very important to measure these parameters on a device level for a precise temperature compensation. In this step, we determine the temperature and strain sensitivities of the FBG using our test rig. 
       \item Compensation of temperature using manual control: In this step, we  increased and decreased the temperature to which the FBG under test is exposed, and then manually released or increased the strain on the fiber to compensate for the shifts in the FBG wavelengths.   
        \item Selection of material and their nominal dimension: Finally, we develop a CAD model to analyse the thermal characteristics of materials and their nominal dimensions; and optimise the design using FEA. 
\end{enumerate} 
In the subsequent sections, we present the experimental realization of the above design methodology. 

\section{Experimental Procedure}
\label{sec:experimentalproc}
In this section we, first, present the experimental setup that are used to realize the steps outlined by the design methodology
and, then discuss the corresponding results obtained in each step before we proceed to the next step.

\par Throughout the experiments we used a single-mode photosensitive fiber, GF1 (here after we will mention only as fiber), which is a product from Nufern/Coherent. Nominal indices in the operating wavelengths for GF1 is as follows:
core refractive index: $1.4489$ @ $\unit[1\,550]{nm}$, clad refractive index: $1.4440$ @ $\unit[1\,550]{nm}$.
Although the information on elastic modulus of GF1 is not readily available from the manufacturing company, literature suggests that the Young's modulus of a photosensitive fiber is very similar to that of an amorphous silica and can vary within 
$\unit[70]{GPa}$ to $\unit[74]{GPa}$.\cite{Rahman2,Antunes2008}

\par
\begin{figure}[ht]
    \centering
    \includegraphics[width=0.8\textwidth]{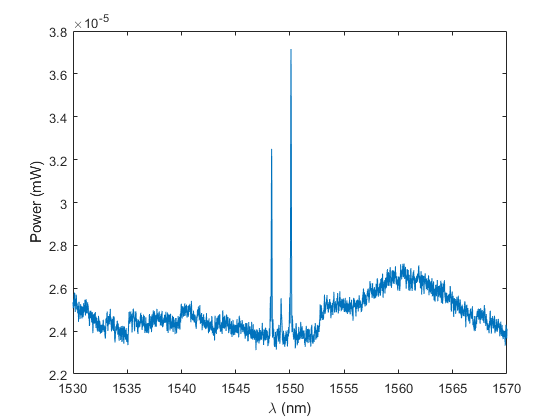}
    \caption{Two FBGs are inscribed on GF1 fiber using phase mask method.}
    \label{fig:FBG}
\end{figure}

 We inscribed uniform FBGs in the core of GF1 with $\unit[244]{nm}$ high power UV laser using phase mask method. Figure~\ref{fig:FBG} shows the spectra (@ $\unit[50]{nm}$ resolution) of the FBGs on GF1, we inscribed two~FBGs, with Bragg wavelengths at $\lambda_{\FBG{}1} = \unit[1548.335]{nm}$, and $\lambda_{\FBG{}2} = \unit[1550.064]{nm}$. 
$\lambda_{\FBG{}2} = \unit[1550.064]{nm}$ is tracked for the temperature compensation tests in our experiments.

\subsection {Development and validation of a custom rig}
\begin{figure}[ht]
    \centering
    \includegraphics[width=0.8\textwidth]{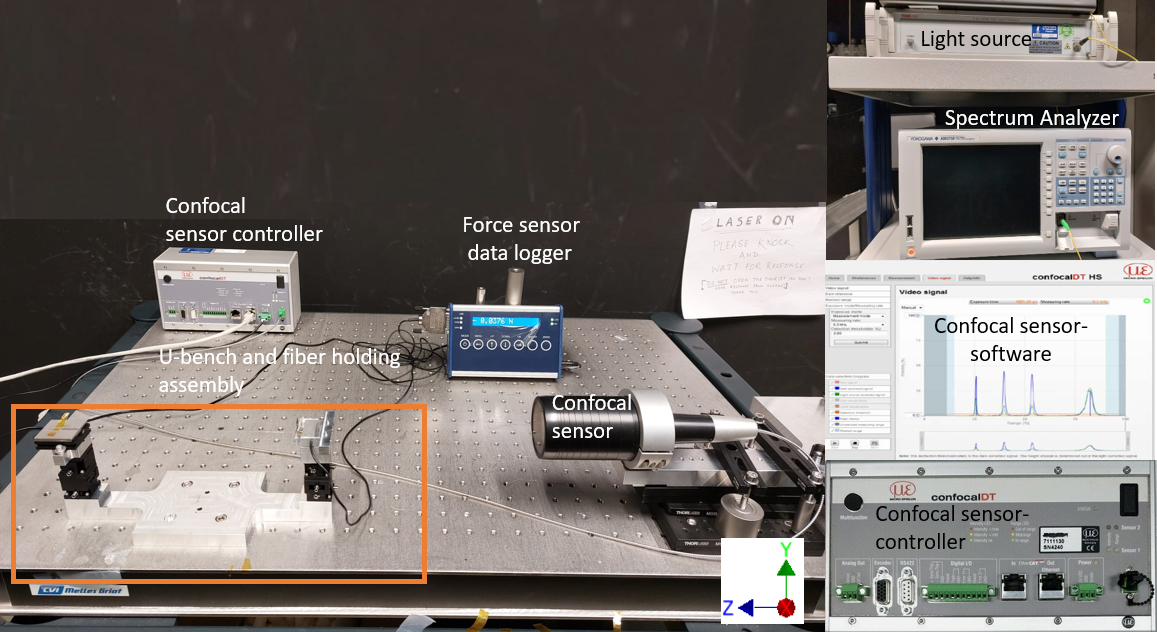}
    \caption{A test bench consisting of a custom-built test rig (U-bench and fiber holding assembly), a confocal sensor and a force sensor for studying several critical aspects of designing an athermal package for temperature compensation in FBG filters used for OH-suppression in astronomy.}
    \label{fig:testrig}
\end{figure}

The test rig comprises of a custom-made U-bench with V-groove fiber holders on its two arms. A series of compact dovetail linear stages with a $\unit[9.5]{mm}$ travel range (M-MT series, mks/Newport) controls the V-groove holder with a precise positioning step of $\unit[1]{\micro m}$.
A thin metal plate is attached to one of the holders, from which the illuminating beam from the confocal distance measuring sensor (Confocal DT IFS2405-28, Micro-Epsilon, resolution $\unit[250]{nm}$) reflects to measure the absolute position, $P$ of the target. This setup measures the elongation or contraction of the anchored fiber. The test bench is shown in Figure~\ref{fig:testrig}. 

\begin{figure}[ht]
    \centering
    \includegraphics[width=0.4\textwidth]{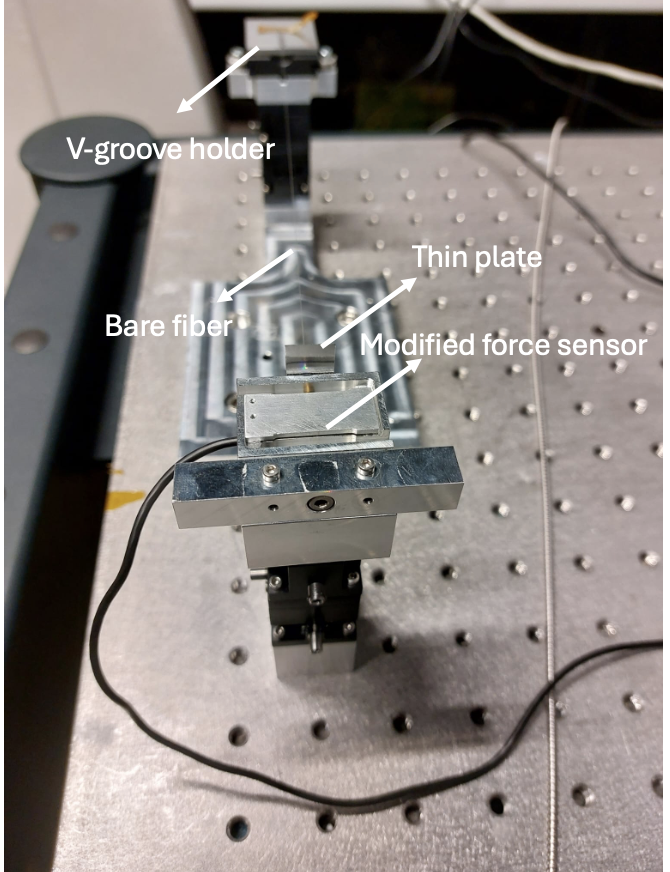}
    \caption{Elastic Modulus test setup}
    \label{fig:EM}
\end{figure}

In this experiment, one of the V-groove holders was replaced with a force sensor (KD34S 2N, ME-Me\ss{}systeme GmbH). The S-arm of the force sensor was modified to hold the fiber at the center of the force point without compromising the sensor mechanism, allowing for the measurement of the optical fiber's Young's Modulus. In this experiment, a bare fiber was used. The fiber was stripped throughout its length to minimise the influence of the acrylate jacket on the Young's Modulus value. The unjacketed fiber was bonded using cyanoacrylate adhesive at one end to the V-groove holder and at the other end to the force sensor arm, as shown in Figure~\ref{fig:EM}. 

\par The length ($L$) of the fiber from one anchored point to the other was measured using  vernier caliper instrument. A force was applied to the fiber along one axis using the dovetail stage. The applied force ($F$) and fiber elongation/contraction ($\Delta l$) were measured using the force sensor data logger and the confocal sensor software, respectively. 
The Young's modulus ($Y$) of the fiber was calculated using the formula $Y=\nicefrac{(F/A)}{(\Delta l/L)}$, where $A$ is the cross-sectional area of the fiber (diameter of the unjacketed fiber is $\unit[125]{\micro m}$). The calculated Young's modulus of the fiber was $\unit[72.29]{GPa}\pm\unit[0.32]{GPa}$, 
which is consistent with the values reported in the literature. Table~\ref{tab:EMTable} in the Appendix, shows the measured data for obtaining the Young's Modulus of the fiber. After the development and validation of the test rig, we proceeded to test the stability of the adhesive material.

\subsection{Test of adhesion stability}
The two ends of the fiber, which included an FBG with a characteristic center wavelength of $\unit[1\,550.064]{nm}$ (measured using the YOKOGAWA \mbox{AQ63758}), were glued to the V-grooves of fiber holders (Thorlabs \mbox{HFV001}) in the test rig using a cyanoacrylate adhesive.
The bonded length was approximately 3 mm long. The adhesion of the optical fibers to the anodized aluminum surface was tested under strain conditions ranging from $\unit[200]{\micro\epsilon}$ to $\unit[1\,500]{\micro\epsilon}$. To ensure temperature compensation across the entire range, stability tests of the stage and glue were conducted at a medium strain of $\unit[550]{\micro\epsilon}$. The test stand was set up on a stable optical bench in a controlled temperature environment of $\unit[292]{K}$.
To maintain the stable temperature of $\unit[292]{K}$ around the fiber, a temperature-controlled breadboard was used within the U-bench test rig, with thermocol insulators isolating the setup from room temperature variations. Additionally, a temperature sensor monitored the temperature of the test setup. All translation stages were locked to enhance stability.
The Bragg wavelength was monitored for 24~hours using an APEX \mbox{AP2683A} optical spectrum analyzer (OSA) with a spectral resolution of $\unit[0.04]{pm}$. Over this period, no shift in the Bragg wavelength was observed, confirming the stability of both the stage and the glue at a medium strain of $\unit[550]{\micro\epsilon}$. Table~\ref{tab:glue_stability} in the Appendix summarizes the results from the adhesion stability test.

\subsection{Selection of pre-strain}
\label{sec:pre-strain}
\begin{itemize}  
\item First, the selection of a suitable pre-strain depends on the stability of the glue. The fiber within the athermal unit undergoes elongation and contraction for temperature compensation, and the glue's stability during these cycles is crucial for selecting the appropriate pre-strain.
\item
The second important factor is the pre-strain allowed during the fabrication process. A small amount of pre-strain is required during FBG fabrication to keep the fiber taut and prevent the formation of tilted gratings. High pre-strain is not advisable, as the fiber tends to slip through the fiber holders during the fabrication process. If the fiber moves during fabrication, the desired wavelength can not be achieved. Additionally, the pre-strain applied to the FBG in the athermal unit at room temperature should match the pre-strain applied during the FBG inscription process to ensure the characteristic center wavelength of the FBG aligns with the desired wavelength of the OH lines.
\item
 Finally, based on the above criteria, an optimum pre-strain is chosen according to the temperature range for which the athermal package is designed to compensate. When the surrounding temperature decreases, the center wavelength of the FBG shifts towards shorter wavelengths due to the dominant effect of the fiber's thermo-optic coefficient ($\zeta$)
over its thermal expansion coefficient ($\alpha_{f}$). This shift must be compensated by the materials of the athermal unit by stretching the fiber and increasing the strain within it. High pre-strain should be avoided to prevent excessive stress on the FBG. Achieving a fine balance in the pre-strain is essential because, when the temperature rises above room temperature, there must be enough room for the fiber to release the strain without bending.
\end{itemize} 

In our application, the temperature range is about $\unit[40]{K}$ (i.e., $\unit[263]{K}$ to $\unit[303]{K}$). Given the typical temperature and strain sensitivities of FBG, approximately \unit[8] to $\unit[10]{\micro\epsilon}$ is required to compensate for $\unit[1]{K}$ temperature variation. From a room temperature of $\unit[295]{K}$ to $\unit[263]{K}$, we must allow an elongation corresponding to a strain of $\unit[240]{\micro}\epsilon$ to $\unit[300]{\micro}\epsilon$ to compensate for temperature variations and a contraction corresponding to about $\unit[150]{\micro}\epsilon$ to compensate for temperature variations from $\unit[263]{K}$ to $\unit[303]{K}$.
In this experiment, the anchored length of the fiber is $L = \unit[121.07]{mm}$ (measured using a vernier caliper), $\lambda_\text{B} = \unit[1\,550.0640]{nm}$ (measured by APEX OSA), and the position, $P$ (measured by the confocal sensor) is $\unit[16.68879]{mm}$. By applying a tension along the axis of the anchored fiber, we observed $\lambda_\text{B}$ has shifted to $\lambda_{\text{B}1} = \unit[1\,550.2680]{nm}$ while $P1 = \unit[16.66540]{mm}$. The strain $\epsilon$ on the fiber is calculated as $(P1 - P)/L = \unit[193]{\micro\epsilon}$, which conforms all the above three important factors for selecting the pre-strain.

Next, we conduct experiments to determine the strain and temperature sensitivities of the FBG.

\subsection{Determination of strain and temperature sensitivities of FBG}
\label{sec:strain_temp}
During the following experiments, large intervals were chosen between the strain and temperature steps to achieve better accuracy. 

\begin{figure}[ht]
{\centering
\includegraphics[width=0.8\textwidth]{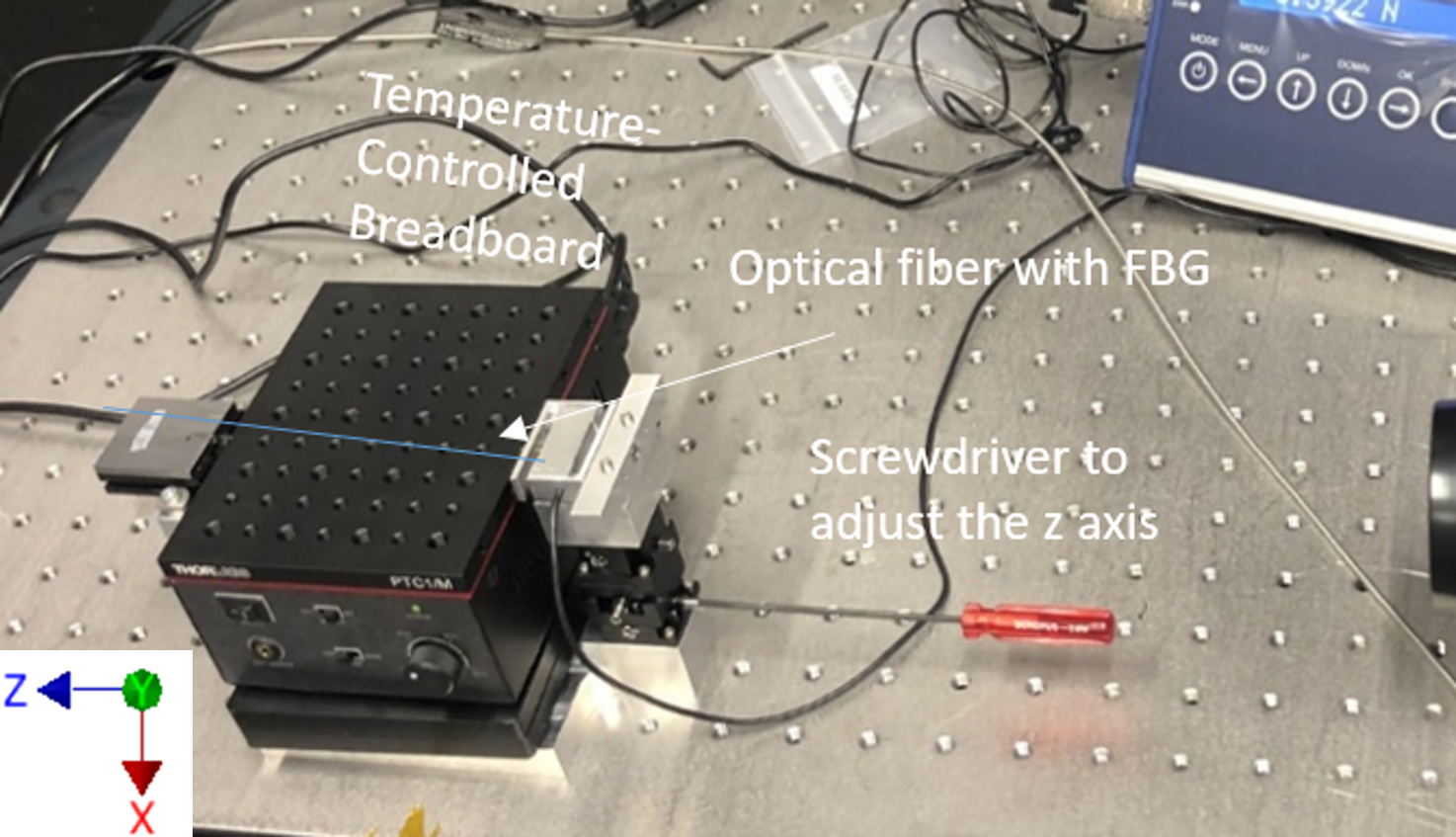}
\caption{Test bench for measuring strain and temperature sensitivities of the FBG.}
\label{fig:system_temp}}
\end{figure}

\begin {itemize} 
\item 
Determination of strain sensitivity: The fiber ends (consisting of the FBG) were bonded onto the \mbox{V-groove} holders in the test rig using the cyanoacrylate glue. A pre-strain was applied, as described in the previous section. One end of the anchored fiber was subjected to an axial strain by translating the micro-translation stage towards an outward axial direction. The position, $P$, of this end of the anchored fiber was measured using the confocal distance measuring sensor. The corresponding shift in Bragg wavelength was measured using the APEX OSA. From the measured data, the strain sensitivity was calculated using $(\lambda_{B1} - \lambda_{B})/(\Delta P/L) = \unit[1.22]{pm/\micro\epsilon} \pm \unit[0.0159]{pm/\micro\epsilon}$, $\Delta P$ is the elongation in the fiber due to the applied axial strain.

\item
Determination of temperature sensitivity:
The unstrained fiber was heated and cooled using the tem\-per\-a\-ture-con\-trolled breadboard \mbox{PTC1/M} (Thorlabs). A change in temperature, $\Delta T$, shifts the reflected wavelength  (measured using Apex OSA) to longer or shorter wavelengths depending on the increase or decrease of the temperature, respectively. Figure~\ref{fig:system_temp} shows the experimental setup for this measurement.
The temperature sensitivity of the FBG was measured using $\left(\lambda_{B1} - \lambda_{B}\right)/\Delta T = \unit[9.77]{pm/K} \pm \unit[0.02]{pm/K}$. Since, the temperature sensitivity of an FBG is different under strained condition, we next, measured the temperature sensitivity of the FBG under the influence of the pre-strain, the temperature sensitivity is found to have slightly enhanced from that of the unstrained FBG. The temperature sensitivity of the FBG under the chosen pre-strain is $\unit[10.24]{pm/K} \pm \unit[0.05]{pm/K}$.
\end{itemize}

Based on the results of the two aforementioned experiments, we have calculated that the FBG under test requires $\approx \unit[8.39]{\micro\epsilon}$ to compensate for $\unit[1]{K}$ temperature variation. This value is crucial and must be measured with high precision to achieve temperature compensation with sub-picometer accuracy. Therefore, cross-verifying the strain required to compensate for temperature variations is essential. Figure~\ref{fig:system_temp} illustrates the U-bench setup with the bonded FBG and the 
temperature controller used in this experiment.

\subsection{Compensation of temperature using manual control}
\label{sec:manualcont}
This experiment directly measures the effect of temperature changes on the FBG. The temperature of the test setup is first increased or decreased, and then the strain on the anchored fiber is manually adjusted to compensate for the shift in the Bragg wavelength. This adjustment is done by translating the fiber holder stage along the fiber axis. We compare the results from this experiment with the previous measurements derived from the temperature and strain sensitivities of the FBG.

To determine the fiber's behavior in varying temperatures, specifically, how much it needs to be pulled when it gets colder or how much pre-strain should be released at higher temperatures, we used the custom test rig and the temperature controller. This allowed us to understand how the compensator must work to maintain the reflected wavelength constantly aligned to that of the desired wavelength.  

The optical fiber was glued to V-grooves on the fiber holders (Thorlabs \mbox{HFV001}) using 
superglue, with the Bragg gratings positioned between the two glued points. The necessary pre-tension was achieved by pulling the adhesive points away from each other in the $z$-direction, i.e., along the fiber axis using the XYZ~stage. The preload setting was verified using the spectrum analyzer. First, the temperature at the test setup was set at $\unit[295]{K}$. Then, the temperature was increased from $\unit[293]{K}$ to $\unit[313]{K}$. The shift in the Bragg wavelength was measured using the APEX OSA. Once the temperature stabilized, the shift in the reflected wavelength was compensated by moving the XYZ stage to apply compression along the axis of the fiber. The change in the movement/position of the stage was measured with the confocal sensor, which allowed us to determine the position change of the stage in the $z$-direction and the necessary fiber length changes for compensation. The experiment was repeated in a similar way when the temperature was decreased from $\unit[313]{K}$ to \unit[283]{K}. By manually adjusting the axial tension, the shift of the Bragg wavelength towards shorter wavelengths was compensated.
We measured that $\approx \unit[8.6]{\micro \epsilon}$
is required to compensate $\unit[1]{K}$ temperature variation, which gives an error of $\unit[2.3]{\%}$ when compared to the previously obtained values from the strain and temperature sensitivities of the FBG from Section~\ref{sec:strain_temp}. This error will introduce an over-compensation or under-compensation of $\unit[0.25]{pm}$ in the characteristic Bragg wavelength. Although this is considered a negligible error, we will discuss potential sources of error and solutions to minimize them in the following sections.

\subsection{Selection of material and their nominal dimensions}
The choice of materials (material 1 and 2 as shown in the Figure \ref{fig:concept}) and the lengths of the components, including the fiber, play a crucial role in achieving high accuracy in the design. Athermalization results from the response to temperature changes of different components made from different materials. Ideally, the changes in the lengths of the materials holding the fiber should completely cancel out the shift in the FBG due to temperature variations. We formulated equations for our design using materials with high and low CTE and obtained the nominal dimensions of the components. With these nominal dimensions, we used a CAD model to optimize the calculated values (from the equations) and the empirical values (related to the FBG) to determine the lengths of the components using finite element analysis (FEA). We compared these optimized results with our calculated results and found an error of $\pm\unit[0.004]{\%}$, amounting to a negligible error ($\ll\unit[1]{pm}$) in the compensation of the FBG filter for \unit[1]{K} temperature variation. In other words, this value suggests that sub-picometer accuracy for the temperature compensation can be achieved in our design. From the FEA results, the absolute value required to compensate for \unit[1]{K} temperature variation is $\approx \unit[8.79]{\micro\epsilon}$. When compared with the experimental data obtained in Section~\ref{sec:manualcont}, an error of $\approx \unit[2]{\%}$ is calculated.  This error will introduce only $\approx \unit[0.23]{pm}$ in under-compensation or over-compensation in the Bragg wavelength for \unit[1]{K} temperature variation. The FEA results also validate that our experimental setup is robust.

As this work focuses on the step-by-step design methodology and emphasizes the critical aspects of the design, the details of the mathematical modeling and the FEA analysis leading to the manufacturing of the prototype are beyond the scope of the present work. We will address the aspects concerning the manufacturing of the prototype in future communications (Part 2: Manufacturing and Testing).

\section{Discussions}
\label{sec:discussions}
\subsection{Error mitigation}
In the design of the athermal unit, we specifically focused on minimizing errors while studying the critical physical properties of the fiber and the FBG. To our knowledge, this is the first time, different approaches to accurately achieving FBG properties have been demonstrated. All experiments were performed in controlled temperatures at the laboratory, using high-precision equipments, and on a stable optical table to which the components were secured firmly with screws to avoid vibrations. The confocal sensor beam was protected to avoid turbulences that could falsify the measurements.
We conducted several measurements to compare the data obtained. The standard deviation of the measurements indicated low variability, ensuring the repeatability of the setup. However, accuracy depends on several factors. We identified the following to be the possible sources of error in our experiments.

\begin{figure}[ht]
    \centering
    \includegraphics[width=0.9\textwidth]{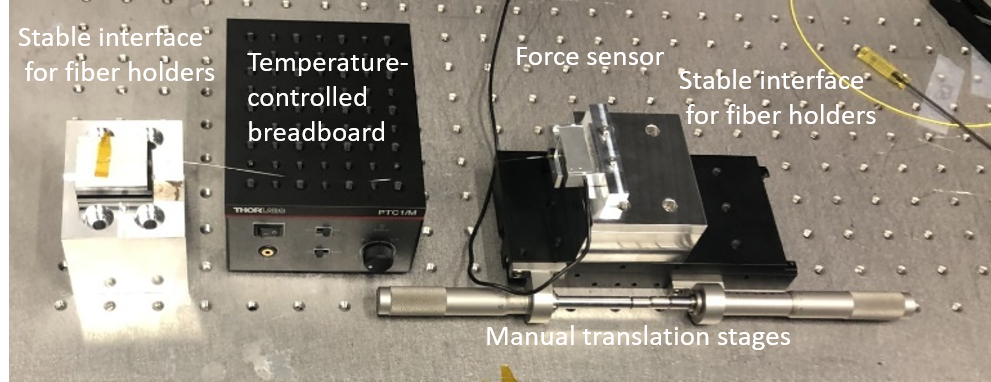}
    \caption{Improved test rig with stable interface for fiber holders}
    \label{fig:UpgradedTestRig}
\end{figure}

\begin{itemize} 
\item Force sensor:  The bending of S-arm of the force sensor under the axial force can reduce the accuracy. For this reason, we used the force sensor only to validate our setup and not for determining a critical parameter, i.e., strain sensitivity of the FBG.
\item XYZ stages: Although the stages were firmly locked during measurements, their structure and arrangement formed a tall and thin column on top of which the fiber was anchored. This arrangement can generate a relatively large moment at the base of the test setup. To minimize this type of error and stay within tolerance, we measured larger distances, e.g., we used a U-bench that holds a longer length of fiber, $\approx \unit[210]{mm}$, instead of the designed length of $\approx \unit[110]{mm}$. The longer fiber length helped achieve better accuracy in our measurements.  
\end{itemize} 
For our future experiments, we have improved the test setup to enhance stability and reduce sources of error, such as tall, thin columnar structures and multiple mini-stages  (Figure~\ref{fig:UpgradedTestRig}). 
While we emphasize the importance of accurately determining the strain and temperature sensitivities of the FBG, we also acknowledge that overall athermalization is achieved only when the materials, dimensions, and other factors are chosen accurately. To minimize further errors, we performed 3D modeling and FEA analysis to optimize the material dimensions in our design. 

It is however, important to note that we are still in the development phase and have not yet calculated the non-linear effects due to temperature changes in the athermal package. These non-linear effects could impact the accuracy of the athermalization (we aim to achieve sub-picometer precision and accuracy).

\subsection{Novel aspects of the design}
Finally, we highlight the following novel aspects of our design:
\begin{itemize}
\item We have discussed the importance of the pre-strain selection in Section~\ref{sec:pre-strain}. Compared to other commercially available packages for FBGs, one pivotal distinction of our athermal unit is its capability to maintain the precise value of the pre-defined strain utilized during filter fabrication. This is a stringent criterion the athermal package must meet, as the FBG filter wavelength must match with OH lines to effectively mitigate their influence on the measured science light from telescopes. Our design includes features for aligning and accurately adjusting the pre-strain to match the pre-strain used during FBG inscription. This aspect becomes especially crucial when employing identical sets of FBG filters in combination with photonic lanterns for OH-suppression in on-sky tests. We intend to elaborate on our methods for achieving pre-strain adjustability in detail in forthcoming communications focusing on the manufacturing and testing of the prototype.

\item Another important feature of our athermal package, which other commercial athermal packages do not offer, is the ability to mount FBG filters up to \unit[210]{mm} long. Additionally, using our developed mathematical models to determine the length of the components, we can calculate the necessary dimensions to mount even longer FBG filters, making our athermal package scalable. Other FBG based devices, e.g., Fabry-P{\'e}rot interferometers can benefit from such athermal unit to maintain its stability at ambient temperatures, especially when the long-length chirped gratings are used to fabricate these in-fiber frequency combs \cite{Madhav}.
\end{itemize}

\section{Conclusions}
\label{sec:conclusions}
In this paper, we present the design and development of an athermal package for stabilizing the FBG filters intended for ground-based NIR telescopes. We used a custom test rig to study the key factors involved in the design of the athermal package. The critical aspects of the design include selecting the optimum fiber pre-tension, studying the fiber-glue adhesion performance, and accurately determining the strain and temperature sensitivities of the FBG. In our study, we implemented a) careful measurement techniques, b) stable test environments, c) re\-dun\-dan\-cy/cross-ver\-i\-fi\-ca\-tion processes, and d) advanced modeling and simulation to enhance the accuracy of the results. The low variability in our measured data ensured the repeatability of our experimental setup. The CAD model of the athermal unit was optimized for its thermal behavior using FEA. The optimized results show an error of 
$\pm \unit[0.004]{\%}$ in the nominal dimensions of the materials, amounting to a negligible error in the compensation of the FBG filter under temperature variations. Based on these encouraging results, we manufactured the prototype, the characterization of which will be communicated in future publications. The step-by-step process of designing a high-precision self-compensating package outlined in this work lays a foundation for broader applications of temperature-stabilized FBG devices in various scientific and industrial fields.

\acknowledgments % equivalent to \section*{ACKNOWLEDGMENTS}
This work is supported by DFG (Deutsche Forschungsgemeinschaft project no.~455425131). We acknowledge the cooperation from Mr.~Thomas Jahn, 3D and Multi Object Spectroscopy, AIP, Mr.~Pa\v{s}ko Roje, Project Management, AIP, for an introduction to the functionalities of the position sensor and we acknowledge the cooperation from Dr.~Dragan Marinkovic, Department of Structural Mechanics and Analysis, Technical University Berlin.

\clearpage%  \todo{check at the end of the editing process, if the clearpage command at this position is still needed. \ho}
\appendix
\section*{APPENDIX}
%%%%%%%%%%%%%%%%%%%% Table 1 %% https://www.tablesgenerator.com/
\begin{table}[ht]
\caption[Young's Modulus of fibre]{The table shows the measured data for obtaining the Young's Modulus ($Y$) of the fiber.}
\label{tab:EMTable}
\begin{center}
\begin{tabular}{lllll}
\hline
$L$ / (mm)                & $P$ / (mm)   & $\Delta l$ / (mm)          & $F$ / (N)                 & $Y$ / (GPa)              \\ \hline \hline
\multirow{2}{*}{$211.22$} & $17.63956$   & \multirow{2}{*}{$0.11979$} & \multirow{2}{*}{$0.5$}  & \multirow{2}{*}{$71.87$} \\
                          & $17.51977$   &                            &                         &                          \\ \hline
\multirow{2}{*}{$211.34$} & $17.51977$   & \multirow{2}{*}{$0.11842$} & \multirow{2}{*}{$0.5$}  & \multirow{2}{*}{$72.75$} \\
                          & $17.63819$   &                            &                         &                          \\ \hline
\multirow{2}{*}{$211.22$} & $17.63821$   & \multirow{2}{*}{$0.14267$} & \multirow{2}{*}{$0.6$}  & \multirow{2}{*}{$72.42$} \\
                          & $17.49554$   &                            &                         &                          \\ \hline
\multirow{2}{*}{$211.38$} & $17.49304$   & \multirow{2}{*}{$0.14575$} & \multirow{2}{*}{$0.61$} & \multirow{2}{*}{$72.13$} \\
                          & $17.63879$   &                            &                         &                          \\ \hline
\end{tabular}
\end{center}
\end{table}
%\begin{figure}[ht]
% \centering
% \includegraphics[width=0.8\textwidth]{EM_Table.png}
% \caption{Measured data for obtaining the Young's Modulus of the fiber}
% \label{tab:EM_Table}
%\end{figure}

%%%%%%%%%%%%%%%%%%%% Table 2 %% https://www.tablesgenerator.com/
\begin{table}[ht]
\caption{The table summarizes the results from the stability test of the adhesive material}
\label{tab:glue_stability}
\begin{center}
\begin{tabular}{p{1.3cm}p{1.5cm}p{1.3cm}p{2.9cm}p{6.5cm}}
\hline
Tem\-per\-a\-ture & Pre-strain  & In\-su\-la\-tion used & Shift in Bragg wave\-length     & In\-fer\-ence                                                                                                                                \\ \hline \hline
$\unit[295]{K}$             & $\unit[1\,500]{\micro}\epsilon$                                        & No              & blue shift: $\unit[83.8]{pm}$ & The glue performance is unstable at high strain.                                                                                         \\
$\unit[295]{K}$             & $\unit[550]{\micro}\epsilon$                                           & No              & red shift: $\unit[18]{pm}$    & Glue performance is better at medium strain, however, the room temperature is influencing the surrounding temperature of the test setup. \\
$\unit[295]{K}$             & $\unit[550]{\micro}\epsilon$                                           & Yes             & red shift: $\unit[0.6]{pm}$   & Glue performance improves, however, the room temperature slightly influences the surrounding temperature of the test setup.              \\
$\unit[292]{K}$             & $\unit[550]{\micro}\epsilon$                                           & Yes             & no shift                      & Glue performance is good; the temperature stability around the test set up is achieved.                                                  \\ \hline
\end{tabular}
\end{center}
\end{table}
%\begin{figure}[ht]
%    \centering
%    \includegraphics[width=0.8\textwidth]{Glue_stability.png}
%    %\caption{Measured data for obtaining the Young's Modulus of the fiber}
%    \label{tab:glue_stabilitytest of the adhesive material}
%\end{figure}

%\todohint[inline]{1. 	Find the Young’s Modulus of the photosensitive fiber: validates the measurement from the experimental test set-up\\
%2. 	Test the stability with glue: overnight test at room temperature\\
%3. 	A test bench in a controlled environment for measuring the strain and temperature response of the FBG used in the test setup\\
%4. 	A manual control of the fiber expansion/contraction to compensate for decrease/increase in temperature\\
%5. 	Error calculation with measured strain-temperature sensitivities of the FBG used in the setup\\
%6. 	Pre-strain selection\\
%7. 	CAD design and ANSYS modeling for material dimension selection that allows the selected pre-strain (This part will be presented in more detail in a future correspondence: part 2: modeling, fabrication, and characterization) FEM IMAGENS\\
%-\ar}

%\label{sec:experimentalsetup}
%\todohint[inline]{Details (CAD and small parts fabrication) about the test bench with the measurement using the confocal sensor.REAL FOTOS OR sketchß}
%\section {Open Questions}
%1 should we writing the description of the hartware i.e spectra analaizer, toleransen of the confocaler sensor etc?
%\appendix

\clearpage
% References
\FloatBarrier
%\todoremark{Note: bibliography data is in report.bib}
\bibliographystyle{spiebib} % makes bibtex use spiebib.bst
\bibliography{report} % bibliography data in report.bib
%
%

%\clearpage\pagestyle{empty}\thispagestyle{empty}\listoftodos\todo{just comment this line out at the end.} %%%%% comment this line out once the document is completed
%
%
%\begin{figure} % https://tex.stackexchange.com/questions/375337/weird-looking-square-root-on-pgfplots
    %\centering
    %\begin{tikzpicture}
        %\begin{axis}[
            %xlabel={$x$-axis},
            %ylabel={$y$-axis},
           % title={Plot from External Data File},
           % grid=both
        %]
            %\addplot[
                %color=blue,
               % smooth, 
                %no markers]
               % table [col sep=space] {temp_Spectrum.txt};
            %\addlegendentry{Data}
        %\end{axis}
    %\end{tikzpicture}
   % \caption{Plot of the data from temp_Spectrum.txt}
    %\label{fig:data-plot}
%\end{figure}
\end{document}